# Hydrogen Activation via Dihydride Formation on a $Rh_1/Fe_3O_4(001)$ Single-Atom Catalyst

Chunlei Wang*‡[1], Panukorn Sombut‡[1], Lena Puntscher[1], Nail Barama[1], Maosheng Hao[2], Florian Kraushofer[1], Jiri Pavelec[1], Matthias Meier[1,3], Florian Libisch[2], Michael Schmid[1], Ulrike Diebold[1], Cesare Franchini[3,4], and Gareth S. Parkinson*[1]

[1]Institute of Applied Physics, TU Wien, Vienna, Austria

[2]Institute of Theoretical Physics, TU Wien, Vienna, Austria

[3]Faculty of Physics, Center for Computational Materials Science, University of Vienna, Vienna, Austria

[4]Dipartimento di Fisica e Astronomia, Università di Bologna, Bologna, Italy

**ABSTRACT**

Hydrogen activation is a key elementary step in catalytic hydrogenation. In heterogeneous catalysis, it usually proceeds through dissociative adsorption on metal nanoparticles followed by surface diffusion or spillover, whereas homogeneous catalysts activate $H_2$ through dihydride or dihydrogen intermediates at a single metal center. Here, we show that isolated Rh adatoms supported on $Fe_3O_4(001)$ activate hydrogen through formation of a stable dihydride species without atomic H spillover. Temperature-programmed desorption, X-ray photoelectron spectroscopy, and scanning tunneling microscopy collectively reveal strong (≈1 eV) hydrogen adsorption exclusively at isolated $Rh_1$ sites, while isotope-exchange experiments further demonstrate that hydrogen remains localized. Density-functional theory based calculations indicate a barrierless conversion from molecular $H_2$ to the dihydride, and random-phase approximation calculations further confirm the relative stability of the dihydride. Together, these results show that single-atom Rh sites cleave and bind $H_2$ through a dihydride pathway analogous to homogeneous complexes, establishing a mechanistic bridge between homogeneous and heterogeneous catalysis.

# 1. Introduction

Catalysis plays a critical role in the modern economy, and tremendous effort is devoted to improving catalytic performance in terms of activity, selectivity, and atomic efficiency.[1] Hydrogenation, the process of adding hydrogen to unsaturated molecules, is a key transformation used in fine chemical synthesis, petroleum refining, and pharmaceutical production.[2] Understanding how catalysts activate $H_2$ is therefore of fundamental importance, as this step often determines both rate and selectivity.[3]

Hydrogenation can be achieved by both heterogeneous and homogeneous catalysts. Industrial processes typically prefer heterogeneous systems because they are robust, easy to separate from products, and reusable. Homogeneous catalysts, by contrast, excel when high selectivity, mild conditions, and precise control over molecular transformations are required.[4] Bridging the advantages of both regimes, by combining molecular precision with practical operability, has long been a central goal in catalysis research.[5]

In homogeneous catalysis, the metal center is coordinated by molecular ligands, and $H_2$ activation is generally achieved by formation of a dihydride or dihydrogen complex in which the H–H bond is cleaved or partially elongated.[6] These intermediates are widely regarded as the key to the high reactivity and selectivity observed in homogeneous catalysis. In contrast, hydrogen activation on supported metal catalysts usually occurs through dissociative adsorption on nanoparticles, followed by migration of atomic hydrogen either across the metal surface or onto the support (hydrogen spillover).[7] Although such mechanisms can enhance overall activity, the resulting mobile hydrogen reservoir often compromises selectivity for partially hydrogenated products.[4] This highlights the need for alternative activation modes on solid catalysts that could emulate the controlled reactivity of molecular systems.

Single-atom catalysts (SACs), in which individual metal atoms are anchored to a support, offer such an alternative.[8] These are heterogeneous catalysts but possess structural similarities to homogeneous systems, particularly in their metal-ligand coordination environment. As a result, SACs have been proposed as a platform to bridge homogeneous and heterogeneous catalysis.[1a, 5c, 9] However, beyond structural resemblance, achieving this bridge requires that SACs operate through similar catalytic mechanisms. $H_2$ activation on SACs is commonly reported to proceed via metal–H formation or through hydrogen spillover onto the supports.[10] For example, Doudin et al. observed hydrogen dissociation and spillover on a $Pd_1/Fe_3O_4(001)$ model SAC at room temperature.[11] More recently, Pacchioni and coworkers calculated that stable dihydride and dihydrogen complexes could also form on SACs, with important implications for the electrochemical hydrogen evolution reaction.[12] However, direct experimental evidence for such species under well-defined conditions has remained elusive

In this work, we experimentally demonstrate that $Rh_1$ species on a reducible $Fe_3O_4(001)$ support strongly adsorb $H_2$ without atomic hydrogen migration and spillover. Instead, a dihydride species forms, analogous to $H_2$ activation in homogeneous catalysis. These findings suggest that heterogeneous SACs can provide mechanisms similar to those of highly selective homogeneous catalysts.

## 2. Results and discussion

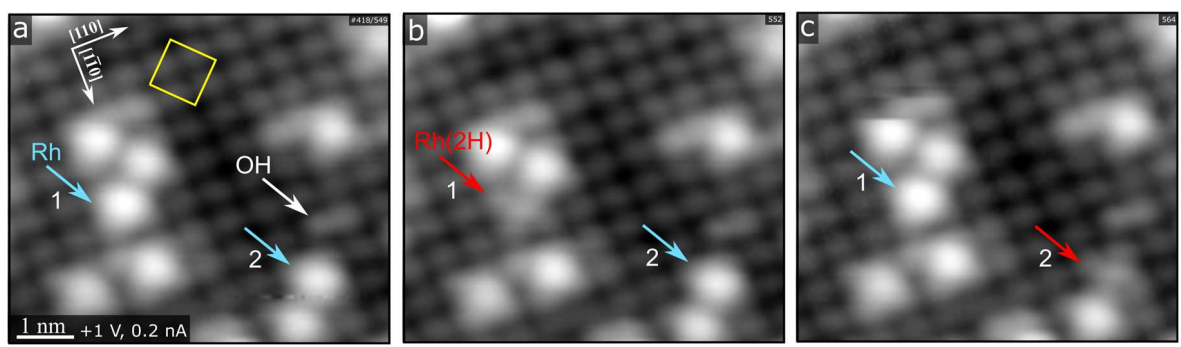

**Figure 1. Time-lapse STM shows reversible $H_2$ adsorption and desorption on 0.2 ML Rh/Fe$_3$O$_4$(001).** STM was conducted at room temperature under a partial pressure of $4 \times 10^{-9}$ mbar $H_2$. The three frames of STM images depict representative states of Rh adatoms during $H_2$ exposure. Sites that adsorb $H_2$ exhibit reduced apparent height (red arrows in b and c). The cyan arrows may represent the Rh$_1$ states either prior to $H_2$ adsorption or following $H_2$ desorption. Some surface hydroxyls were already formed on Fe$_3$O$_4$(001) during the initial sample preparation and are marked with a white arrow. The yellow square in (a) marks the reconstructed surface unit cell of Fe$_3$O$_4$(001).

The Fe$_3$O$_4$(001) surface reconstructs into a ($\sqrt{2} \times \sqrt{2}$)R45° pattern when prepared under UHV (Figure 1a, yellow square). This structure arises from an ordered arrangement of subsurface cation vacancies and interstitials, resulting in an oxidized surface layer.[13] The as-prepared surface contains several defects, most notably hydroxyl species marked by white arrows in Figures 1a and S1a. These originate from dissociative adsorption of trace water at oxygen vacancies created during sputtering and annealing. Although hydroxyls are not directly visible in STM, they modify the contrast of nearby Fe atoms, which appear brighter in empty-state images.[14] This assignment has been confirmed by direct deposition of atomic hydrogen.[15] At room temperature, the hydroxyls hop between two equivalent oxygen sites within a surface unit cell, producing a characteristic back-and-forth motion easily recognized in STM movies.[14a] Apart from the hopping, the number and position of the hydroxyls remain unchanged throughout the following measurements.

For STM measurements, 0.2 monolayers (ML) of Rh were deposited onto the as-prepared Fe$_3$O$_4$(001) surface at room temperature. Individual Rh adatoms are marked by cyan arrows in Figure 1. They appear as bright protrusions located midway between surface Fe rows, with a nearest-neighbor distance of 8.34 Å. The Rh adatoms bind to two surface oxygen atoms on opposite rows (see Figure S1(b-d), consistent with the behavior of other metals on this surface).[16] Additionally, the 2-fold Rh–O coordination observed on Fe$_3$O$_4$(001) is stable under reactive environments,[16a, 17] including high-pressure CO exposure.[18] Time-lapse STM in $4 \times 10^{-9}$ mbar $H_2$ reveals reversible height changes at individual Rh$_1$ sites (Figure 1b,c). The apparent height of a Rh$_1$ adatom is typically 140–160 pm but transiently decreases to 40–70 pm during $H_2$ adsorption. These fluctuations occur randomly and infrequently, indicating reversible $H_2$ adsorption at room temperature. No new bright features or hydroxyls appear near the affected sites, confirming that hydrogen remains localized at Rh$_1$ and does not spill over onto the oxide lattice. An STM movie covering a longer time span and larger area than Figure 1 is provided in the supporting information (Movie S1).

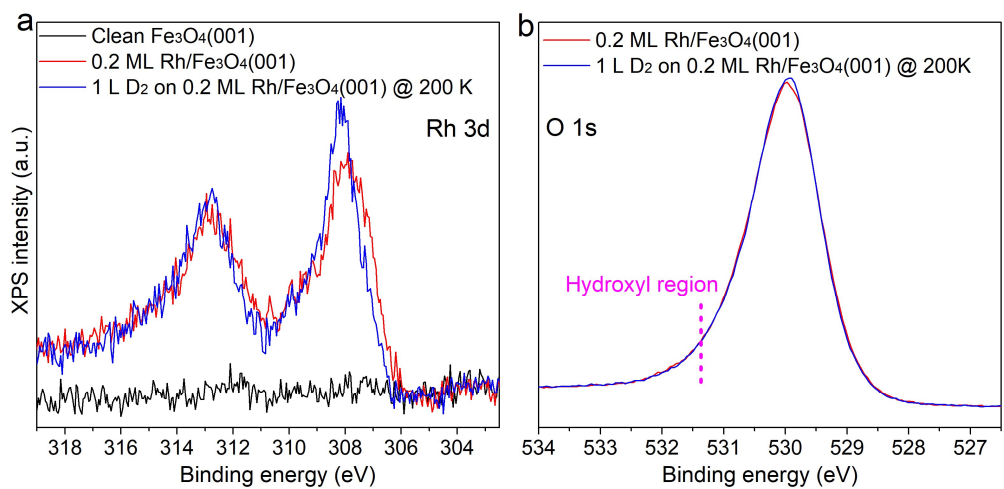

**Figure 2. XPS confirms lack of hydroxyl formation**. XPS spectra of the (a) Rh 3d and (b) O 1s regions collected from clean $Fe_3O_4(001)$ and from 0.2 ML Rh on $Fe_3O_4(001)$ before and after 1 L $D_2$ exposure at 200 K. 1 L is defined as $1.33 \times 10^{-6}$ mbar·s. The pink dashed line in (b) marks the OD region in the O 1s spectrum.

XPS spectra of the Rh 3d and O 1s regions were recorded for 0.2 ML Rh/$Fe_3O_4(001)$ before and after $D_2$ exposure. After adsorption at 200 K, the Rh 3d peak sharpens and shifts slightly to higher binding energy (Figure 2a), indicating modification of the $Rh_1$ electronic state. In the O 1s region, no new feature appears near 531.3 eV where hydroxyl species are expected on $Fe_3O_4(001)$[19], even after prolonged $D_2$ exposure (10 L at 250 K, Figure S2). The absence of this signal confirms that hydrogen remains bound to $Rh_1$ and does not migrate onto the oxide lattice.

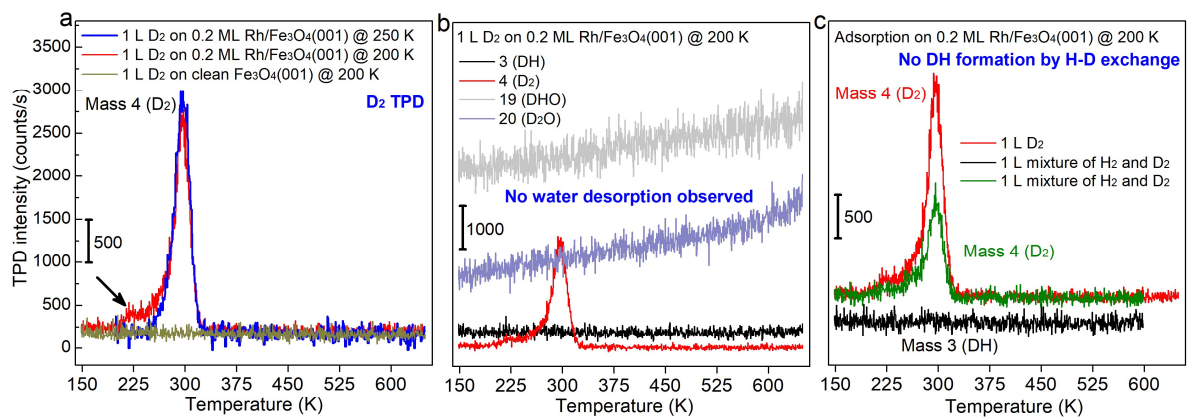

**Figure 3**. **TPD isotope exchange experiments.** (a) TPD spectra acquired after exposing 0.2 ML Rh/$Fe_3O_4(001)$ to 1 L $D_2$ at 250 K and 200 K. A reference spectrum was also recorded after exposing clean $Fe_3O_4(001)$ to 1 L $D_2$ at 200 K. Only the mass 4 ($D_2$) signal is shown. The small arrow indicated stems from the sample holder (see Figures S4-S5). (b) Representative TPD data collected after 1 L $D_2$ exposure on 0.2 ML Rh/$Fe_3O_4(001)$ at 200 K, showing signals for mass 3 (DH), 4 ($D_2$), 19 (DHO), and 20 ($D_2O$). (c) TPD after dosing pure $D_2$ and an equimolar 1 L mixture of $D_2$ and $H_2$ (1:1 partial pressure ratio) at 200 K. No isotope scrambling was observed. In all TPD experiments, the sample was cooled by 50 K after dosing to achieve a perfectly linear temperature ramp in the region where desorption occurs. Curves in (c) are vertically offset for clarity.

To quantify the strength of hydrogen adsorption on $Rh/Fe_3O_4(001)$, TPD experiments were performed using $D_2$ to avoid interference from residual $H_2$ in the vacuum system. The clean $Fe_3O_4(001)$ surface shows no detectable $D_2$ desorption after 1 L exposure at 200 K (Figure 3a, olive trace), confirming that molecular hydrogen binds only weakly to the oxide as on other oxide surfaces[20]. After deposition of 0.2 ML $Rh_1$ and exposure to 1 L $D_2$ at 250 K, a distinct $D_2$ desorption peak appears at 295 K (Figure 3a, blue trace), consistent with the transient adsorption events observed by STM (Figure 1).

Lowering the dosing temperature to 200 K (Figure 3a, red trace) or 150 K (Figure S3, pink trace) does not reveal additional adsorption states. A weak shoulder near 225 K is identified as an artifact from $D_2$ adsorption on the Ta sample holder (Figures S4–S5). Analysis of the main desorption peak using a recently introduced method based on equilibrium thermodynamics[21] (Figure S6) yields an adsorption energy of $0.98 \pm 0.06$ eV, indicating strong but reversible binding of hydrogen to $Rh_1$ sites.

No $D_2O$ desorption is detected up to 600 K (Figure 3b), ruling out formation of OD species on the support (recombinative water desorption occurs at $\approx 550$ K on $Fe_3O_4(001)$).[11] Even at higher doses (5–10 L) or elevated exposure temperatures (200–295 K), no spillover products appear (Figures S7–S8). Signals corresponding to DH or DHO are also absent, excluding isotope scrambling or recombination with residual $H_2$.

Exposure of $Rh_1/Fe_3O_4(001)$ to an equimolar $H_2/D_2$ mixture provides a further test for hydrogen mobility. The $D_2$ signal decreases by roughly half relative to pure $D_2$ dosing (Figure 3c), as expected for competitive adsorption with $H_2$, but no DH signal is observed. Thus, hydrogen adsorbs and desorbs exclusively at isolated $Rh_1$ sites without spillover or isotope exchange. The $H_2$ signal is not shown due to the comparably high background in UHV chamber, hiding the $H_2$ peak in the noise, see Figure S9 for details.

Although individual Rh adatoms occasionally appear in close proximity in STM images, the minimum Rh–Rh separation on $Fe_3O_4(001)$ is sufficiently large to preclude direct Rh–Rh bonding. Importantly, hydrogen adsorption and desorption are characterized by a single, well-defined state in TPD and isotope-exchange experiments, indicating that $H_2$ activation occurs at a unique $Rh_1$ site. No additional hydrogen-related features attributable to Rh–Rh ensembles are observed, demonstrating that the dihydride species identified here is intrinsic to isolated Rh atoms rather than to clustered sites.

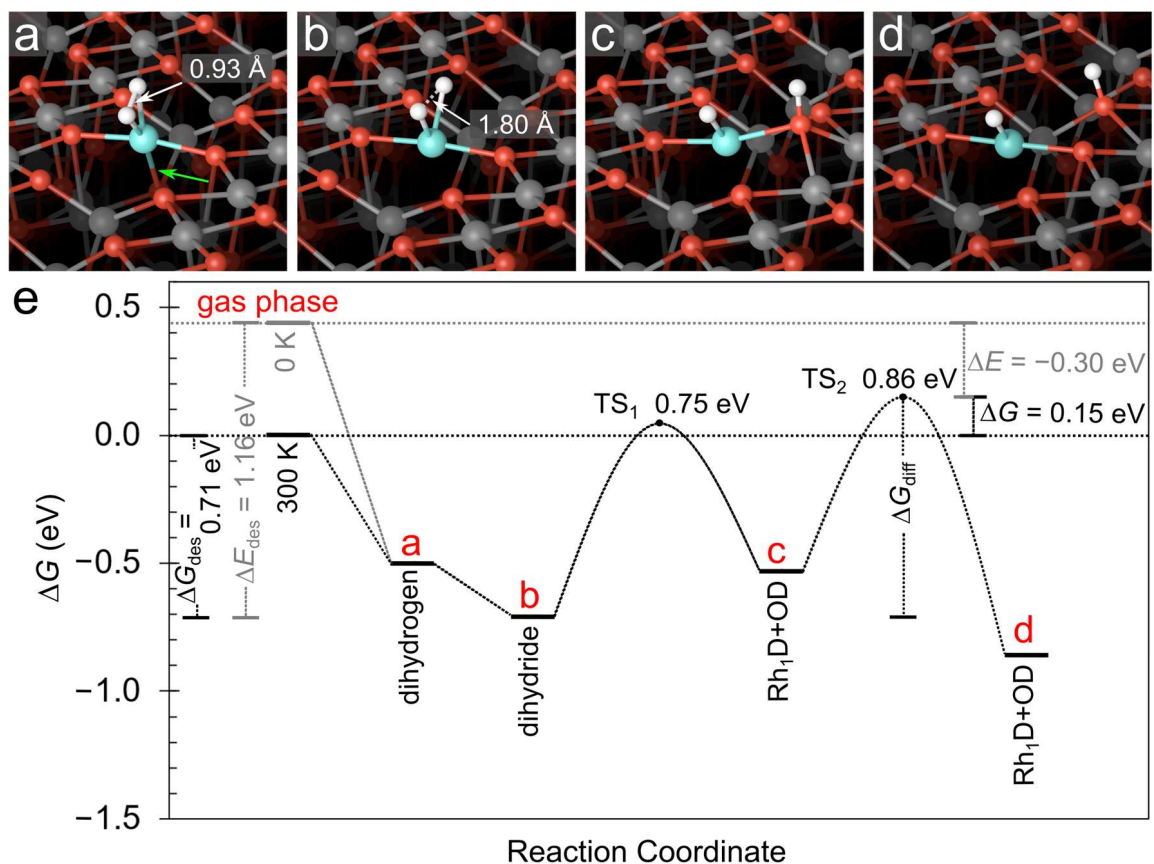

**Figure 4**. **DFT-calculated mechanism of $D_2$ dissociation and spillover on $Rh_1/Fe_3O_4(001)$.** Perspective views are shown for (a) a dihydrogen configuration and (b) a dihydride configuration. In both cases, the $Rh_1$ species is bond to two surface oxygen atoms. An additional weak bond from the Rh to a subsurface O atom (roughly 0.3 Å shorter than in the Rh2D configuration) is indicated by the green arrow. (c-d) Atomic structures of atomic D spillover along the $Fe_3O_4(001)$ surface. Fe atoms are shown in dark gray, and oxygen atoms are red. (e) Computed reaction energy profile for molecular $D_2$ adsorption, dissociation, and hydrogen spillover from a $Rh_1$ site onto the $Fe_3O_4(001)$ substrate. The reported Gibbs free energy ($\Delta G$) values incorporate the entropy difference between the adsorbed state ($S_{ads} = 0$) and the gas phase ($S_{gas}$) (as referenced in database).[22] These $\Delta G$ values are presented for the experimental desorption temperature (300 K) and derived from DFT calculations that include ZPE corrections. The overall reaction mechanism remains identical to the 0 K pathway, except the desorption step, for which the 0 K desorption energy ($\Delta E_{ads}$) is provided for comparative reference. Transition states ($TS_1$ and $TS_2$) and the corresponding activation barriers with respect to the dihydride state (b) are indicated.

Density functional theory (DFT+$U$) calculations were performed to clarify the nature of deuterium adsorption on $Rh_1/Fe_3O_4(001)$. In the following, all values include zero point energy (ZPE) corrections for $D_2$, with corresponding results for $H_2$ shown in the SI. Two local minima were identified (Figure 4a-b). The weak-binding configuration is a dihydrogen-like complex with the D–D distance of 0.93 Å and an adsorption energy of −0.94 eV. In this case, the Rh atom relaxes slightly toward a subsurface oxygen (green arrow in Figure 4a), forming a pseudo-square-planar geometry similar to that found for CO and $C_2H_4$ adsorption on $Rh_1/Fe_3O_4(001)$.[17, 23]

The second, more strongly bound structure corresponds to a dihydride configuration, where the D–D distance is 1.8 Å and the adsorption energy is −1.16 eV. The two deuterium atoms occupy positions above the Rh center, giving the metal a tetrahedral coordination geometry (Figure 4b). This contrasts with the square-planar configuration observed previously for gem-dicarbonyl species on the same surface.[23] This is consistent with the classical ligand-field expectations: weak-field hydride ligands favor tetrahedral coordination, whereas strong-field CO ligands stabilize a square-planar arrangement.[24]

To address the sensitivity of SACs to self-interaction errors, we benchmarked hydrogen binding energetics using the r²SCAN+U and HSE06 functionals, as well as the random phase approximation (RPA) method. The identical energetic ordering obtained across all methods (see Table S1) corroborates the robustness of our results. The dihydride state is lower in energy by 0.22 eV, and CI-NEB calculations suggest a barrierless conversion from the molecularly adsorbed $D_2$ (dihydrogen) precursor to the dihydride product. This transition closely parallels activation mechanisms proposed for homogeneous transition-metal complexes.[6a] In principle, advanced infrared spectroscopy on SACs can provide direct experimental characterization of hydrogen complexes through M–H vibrational modes, as has been reported.[25] However, In our case, the expected IR signals are severely limited by the low loading of metal single atoms and the weak IR activity of hydrogen species.[26]

DFT+$U$ calculations suggest that two $D_2$ molecules could, in principle, adsorb on a single $Rh_1$ atom, with a combined adsorption energy of −1.62 eV (−0.81 eV per $D_2$; Figure S10). Experimentally, however, only one desorption state is observed, and H–D exchange experiments show no isotope scrambling. The absence of a second adsorption state indicates kinetic hindrance, as accommodating a second molecule would require substantial distortion of the Rh center. Similar geometric constraints were previously found for CO adsorption, where binding an additional ligand necessitates vertical displacement of Rh and lateral rearrangement of the first adsorbate. Due to the energetic cost, this is a rare process that is unlikely to coincide with adsorption events under UHV conditions.[23]

We have also considered hydrogen adsorption on substitutional (5-fold coordinated) Rh atoms, which can be also present in this surface (especially after annealing to higher temperatures than used here; see also Figure S1b). DFT+U shows that hydrogen adsorption is weak (−0.39 eV and −0.44 eV for for $H_2$ and $D_2$, respectively) and dissociation (dihydride formation) does not occur at these sites. This agrees with the notion that dihydride formation requires undercoordinated metal atoms.[12]

To understand the lack of spillover at room temperature in the experiments, entropic contributions to the free energy ΔG must be taken into account. Figure 4e compares the free-energy profiles for hydrogen desorption and for hydrogen migration from Rh to a neighboring surface oxygen. Spillover would proceed via an initial hydrogen transfer from $Rh_1$ to a neighboring oxygen atom, followed by diffusion away from the metal center (Figures 4c,d).

The calculated activation barrier for hydrogen migration from $Rh_1$ to a neighboring surface oxygen and subsequent diffusion away from the metal center is 0.86 eV (Figures 4c,d). In contrast, when entropic contributions at 300 K are included, the free-energy barrier for recombinative $H_2$ desorption is 0.71 eV (Figure 4e). Because the spillover barrier exceeds the desorption barrier, hydrogen desorption is kinetically preferred under the experimental conditions, explaining the absence of spillover in experiment.

At first glance, the absence of hydrogen spillover may seem counterintuitive given that the final spillover state is thermodynamically accessible, and more favorable than desorption when disregarding entropy. However, spillover and desorption are fundamentally different processes. Spillover requires a localized, rearrangement at the Rh–O interface. Compared to desorption, it is entropically disfavored because both hydrogen atoms remain surface-bound. In contrast, desorption releases $H_2$ into the gas phase, where the large translational entropy strongly stabilizes the transition state. Kinetics are governed by free-energy barriers rather than the mere DFT-calculated energy barriers. Until desorption, this effectively confines hydrogen to the $Rh_1$ site under the conditions studied here. The corresponding free-energy profile for $H_2$ is provided in Figure S11.

The $\Delta G$ given in Figure 4e adopt the commonly used approximation that the entropy contribution of adsorbed species is negligible ($S_{ads*} \approx 0$), which is generally considered reasonable for small adsorbates such as hydrogen.[27] In our case, the relevant entropy contributions in the adsorbed state are (i) vibrations of the adsorbed deuterium and (ii) the change of the Rh vibrations associated to its transition from a 2-fold to a pseudo-square-planar geometry. The vibration frequencies are given in Table S2. The effect of the deuterium vibrations on adsorption is small (disfavoring desorption by 0.024 eV at 300 K; less for hydrogen). The entropy of the $Rh_1$ adatom slightly decreases upon adsorption (increased vibration frequencies), making desorption more favorable by 0.010 eV. These effects are much smaller than typical DFT errors. Due to the low mass (high vibration frequencies) of H and D, we can also neglect entropy changes upon diffusion to the oxide, which would modify the energy landscape of the spillover process. What remains is the large entropy contribution of the gas phase, which strongly favors desorption in spite of an overall energy gain in the final state of spillover (Fig. 4e).

One particularly revealing aspect of the computed energy landscape is the presence of a local minimum in which one deuterium atom transfers to the oxygen atom directly bonded to Rh (Figure 4c). This configuration lies 0.18 eV above the dihydride minimum, corresponding to an equilibrium population of $\approx 9 \times 10^{-4}$ at 300 K. Although too sparse to be detected by STM, XPS, or TPD, such thermally accessible excitations imply that transient heterolytic D cleavage can occur within the Rh–O ensemble. The calculated activation barrier for this intrapair D transfer (0.75 eV) is slightly lower than that for full spillover (0.86 eV), suggesting that deuterium can shuttle locally between Rh and the neighbouring oxygen without diffusing away. Importantly, this transient hydrogen shuttling remains confined to the immediate Rh–O coordination sphere and does not constitute hydrogen spillover in the classical sense. Rather than diffusive migration across the oxide surface, this local metal–support cooperation temporarily frees a coordination site on Rh, enabling adsorption and reaction of unsaturated molecules directly at the $Rh_1$ site. Hydrogenation can therefore proceed via localized hydrogen transfer from the dihydride without requiring long-range hydrogen transport. This process therefore constitutes metal–ligand-type cooperation confined to a single site, rather than hydrogen (deuterium) spillover in the classical sense[7a], providing a clear mechanistic distinction between local heterolytic activation and long-range D migration across the surface.

The resulting "split-hydride" configuration closely parallels bifunctional $H_2$ activation in molecular complexes, where the H–H bond is cleaved heterolytically across a metal–ligand pair.[6] In this context, the Rh–O pair acts as a functional analogue of a metal–ligand site capable of transient H storage and retrieval. This behaviour represents a key step toward replicating the controlled, site-specific reactivity characteristic of homogeneous hydrogenation catalysts.

More broadly, the combined experimental and theoretical results demonstrate that isolated Rh atoms on $Fe_3O_4(001)$ strongly bind and activate $H_2$ without forming a mobile hydrogen reservoir. The stable dihydride configuration identified here experimentally agrees with recent theoretical predictions that SACs can host discrete $M–H_2$ or metal-dihydride species.[12] By confining hydrogen to well-defined metal-hydride species, the system suppresses spillover and prevents the excessive hydrogen availability that typically causes over-hydrogenation on nanoparticle catalysts. This localization is particularly relevant for selective hydrogenation reactions such as the conversion of acetylene to ethylene, where controlled hydrogen delivery is essential for maintaining high selectivity.[28] Comparable control has been reported in other single-atom hydrogenation catalysts,[9b, 10a] supporting the view that geometric confinement and the covalent metal–oxide interaction stabilize molecular-like hydride species while retaining the robustness of a solid surface. These findings thus provide direct mechanistic evidence that SACs can bridge the conceptual gap between homogeneous and heterogeneous hydrogenation.

## 3. Conclusions

Our combined experimental and theoretical analysis demonstrates that isolated Rh atoms on $Fe_3O_4(001)$ activate hydrogen through the formation of kinetically stabilized dihydride species, without hydrogen spillover to the support. This behavior mirrors the $H_2$ activation mechanism of homogeneous catalysts, establishing direct mechanistic correspondence between molecular and supported single-site systems. The findings demonstrate that isolated single atoms can host discrete metal–hydride species at finite temperature, which is required for enabling selective hydrogenation pathways inaccessible to extended metal surfaces. By translating homogeneous dihydride formation to a solid platform, this work provides a foundation for designing SACs that unite molecular-level precision with the robustness and scalability of heterogeneous catalysis.

**Author information**

*Corresponding email: parkinson@iap.tuwien.ac.at and wangc@iap.tuwien.ac.at

‡ These authors contributed equally to this work

**Notes**

The authors declare no competing financial interests.

**Data availability statement**

The data that support the findings of this study are available from the corresponding author upon reasonable request.

**Acknowledgements**

Funding from the European Research Council (ERC) under the European Union's Horizon 2020 research and innovation programme (grant agreement No. [864628], Consolidator Research Grant 'E-SAC') is acknowledged. This research was funded in part by the Austrian Science Fund (FWF) 10.55776/F81 and the Cluster of Excellence MECS (10.55776/COE5). Chunlei Wang gratefully acknowledges financial support from the FWF through project number 10.55776/PAT1934924. Matthias Meier gratefully acknowledges financial support from the FWF through project number 10.55776/PAT2176923. The Austrian Scientific Computing was used to obtain the computational results. For open access purposes, the authors have applied a CC BY public copyright license to any author accepted manuscript version arising from this submission.

**Keywords**: Scanning tunneling microscopy, single-atom catalysis, density functional theory, metal-oxide surfaces

## TOC graphic

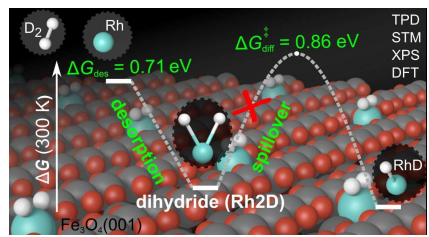

Deuterium (hydrogen) adsorption on single-atom $Rh_1/Fe_3O_4(001)$ catalyst does not induce D (H) spillover onto the reducible $Fe_3O_4$ support. Instead, a $Rh_1$-dihydride (Rh−2D or Rh−2H) species forms and desorbs during annealing.

Supplementary Information:

# Hydrogen Activation via Dihydride Formation on a Rh$_1$/Fe$_3$O$_4$(001) Single-Atom Catalyst

Chunlei Wang[1], Panukorn Sombut[1], Lena Puntscher[1], Nail Barama[1], Maosheng Hao[2], Florian Kraushofer[1], Jiri Pavelec[1], Matthias Meier[1,3], Florian Libisch[2], Michael Schmid[1], Ulrike Diebold[1], Cesare Franchini[3,4], and Gareth S. Parkinson[1]

[1]Institute of Applied Physics, TU Wien, Vienna, Austria

[2]Institute of Theoretical Physics, TU Wien, Vienna, Austria

[3]Faculty of Physics, Center for Computational Materials Science, University of Vienna, Vienna, Austria

[4]Dipartimento di Fisica e Astronomia, Università di Bologna, Bologna, Italy

**The supplementary material includes detailed descriptions of both experimental and computational methods, as well as additional results:**

1) Additional STM Data

2) Additional XPS data

3) Additional TPD data

    a. Origin of shoulder in D$_2$ TPD traces

    b. Deriving desorption energies from TPD

    c. Variation of D$_2$ dosages and adsorption temperatures

    d. H−D isotope exchange experiments

4) Additional computational results

    a. Comparison of adsorption energies with various computational approximations

    b. Higher H$_2$ loading

    c. Estimate of kinetics for hydrogen desorption vs spillover

**Other Supplementary Material for this manuscript includes the following:**
**Movie S1**: Time-lapse STM of 0.2 ML Rh/Fe$_3$O$_4$(001) surface under exposure to 4 × 10$^{-9}$ mbar H$_2$ at room temperature.

**Experimental and Computational Methods**

**Experimental**

Experiments were performed on natural $Fe_3O_4(001)$ (6×6×1 mm) single crystals purchased from SurfaceNet GmbH. Two separate ultra-high vacuum (UHV) systems were used for scanning tunneling microscopy (STM) imaging and for temperature-programmed desorption (TPD)/X-ray photoelectron spectroscopy (XPS) experiments, respectively. Samples were cleaned by repeated cycles of sputtering and annealing in UHV. Sputtering was performed using 1 keV $Ar^+$ ions (STM chamber) or $Ne^+$ ions (TPD/XPS chamber) for 10 minutes, followed by annealing at 900 K for 10 minutes. Annealing in oxygen was used to reoxidize the surface after several cleaning cycles. A final annealing step in $5 \times 10^{-7}$ mbar $O_2$ at 900 K results in the formation of the reconstructed $(\sqrt{2} \times \sqrt{2})R45°$ surface.[1] Rh atoms were deposited using an electron-beam evaporator (FOCUS), and the flux was calibrated with a temperature-stabilized quartz microbalance. The coverage of one monolayer (ML) is defined as one Rh atom per $Fe_3O_4(001)$-$(\sqrt{2}\times\sqrt{2})R45°$ surface unit cell (area density of $1.42 \times 10^{14}$ $cm^{-2}$).

In the TPD chamber, the $Fe_3O_4(001)$ sample was mounted on a Ta backplate, with a thin gold sheet inserted in between to improve thermal contact. The sample was cooled using a liquid-He flow cryostat and heated resistively via the Ta backplate. The chamber includes a home-built molecular beam source that delivers reactants with a calibrated flux (equivalent to an impingement rate of a room-temperature gas at a pressure of $2.66 \times 10^{-8}$ mbar) and a uniform top-hat profile across a 3.5 mm diameter spot on the sample.[2] Gas exposure is given in Langmuirs (L), where 1 L is defined as $1.33 \times 10^{-6}$ mbar·s. A quadrupole mass spectrometer (Hiden HAL 3F PIC) was used for TPD experiments. Deuterium ($D_2$) was employed instead of $H_2$ to avoid background interference from residual $H_2$ in the vacuum chamber. TPD spectra were acquired using a temperature ramp of 1 K/s. The chamber was also equipped with a monochromatic Al/Ag twin-anode X-ray source (Specs XR50 M, FOCUS 500) and a hemispherical analyzer (Specs Phoibos 150) for XPS measurements. A grazing angle of ≈71° with respect to surface normal was used to collect the XPS spectra.

STM studies were conducted using an Omicron μ-STM operated in constant-current mode at room temperature, with an electrochemically etched W tip. A positive sample bias was used in all cases, enabling measurement of the unoccupied electronic states. Sample preparation and Rh deposition were performed in an adjacent chamber. $H_2$ was introduced directly into the analysis chamber for STM time-lapse imaging experiments. The analysis chamber was also equipped with a non-monochromated Al Kα X-ray source and a SPECS Phoibos 100 analyzer for XPS measurements.

**Computational Details**

All calculations were performed using the Vienna *ab initio* simulation package (VASP).[3] The projector augmented-wave (PAW) method[4] was used to describe the near-core regions, and a plane-wave energy cutoff of 550 eV was used. The generalized gradient approximation (GGA) in the form of the Perdew-Burke-Ernzerhof (PBE) functional[5] was used to treat electronic exchange and correlation effects. Dispersion corrections were included following the D3 scheme with Becke-Johnson damping.[6] To adequately describe the strongly correlated Fe 3d electrons, an effective on-side Coulomb interaction $U_{eff}$ of 3.61 eV was applied.[7] The electronic energy convergence criterion was set to $10^{-6}$ eV, and ionic relaxation was performed until the forces acting on ions became smaller than 0.02 eV/Å. The magnetite ($Fe_3O_4$) system was modelled using the experimentally reported lattice parameter of $a$ = 8.396 Å. The asymmetric slab consisted of 7 octahedral Fe and 6 tetrahedral Fe layers. During relaxation, only the topmost 4 layers were allowed to relax, while the bottom 9 layers remained fixed at their bulk positions. Calculations employed a $(2\sqrt{2} \times 2\sqrt{2})R45°$ surface supercell, and Brillouin zone sampling was restricted to the Γ-point only due to the large cell size. A vacuum spacing of 14 Å was included to prevent interaction between periodic images. To account for the small mass of H, zero-point energy (ZPE) corrections were explicitly included in adsorption energy calculations. The adsorption energy ($\Delta E_{ads}$) was computed according to the formula:

$$\Delta E_{ads} = \left(E_{Rh/Fe_3O_4+nH_2} - \left(E_{Rh/Fe_3O_4} + nE_{H_2}\right)\right) + \Delta E_{ZPE}$$

where $E_{Rh/Fe_3O_4+nH_2}$ represents the total energy of the Rh-adatom-decorated $Fe_3O_4$(001) surface with $n$ adsorbed $H_2$ molecules, $E_{Rh/Fe_3O_4}$ denotes the total energy of the Rh-decorated surface without any adsorbates, and $E_{H_2}$ corresponds to the energy of an isolated $H_2$ molecule in the gas phase. The zero-point energy correction ($\Delta E_{ZPE}$) was obtained by calculating vibrational frequencies using the finite-difference method. Only the adsorbed hydrogen atoms, the Rh center, and the oxygen atoms bonded to the hydrogen along the proposed diffusion pathway were allowed to be displaced, while all the other atoms were kept fixed at their optimized positions.

The Gibbs free energy at $T$ = 300 K was then calculated as:

$$\Delta G_{ads} = \Delta E_{ads} - T\Delta S$$

where $\Delta S$ is the entropy change between the adsorbed species (taken as $S$ = 0) and the gas phase.[8] Diffusion activation energies were calculated using the climbing-image nudged elastic band (CI-NEB)[9] method. The electronic self-consistency criterion was set to $10^{-6}$ eV, and the forces on atoms in each image were relaxed to below 0.05 eV/Å. Additional calculations were performed using the regularized-restored strongly constrained and appropriately normed meta-generalized gradient approximation (r$^2$SCAN) [10] with the $U_{\text{eff}}$ of 3.10 eV[11] and the hybrid functional HSE06[12] to validate results and assess the accuracy of the employed methods. To further assess the effects of electronic correlation, we also performed additional random-phase approximation (RPA) calculations based on the hybrid DFT orbitals, using the more accurate PAW projectors provided with VASP to better capture the local electronic structure around the Rh atom.[13] Due to computational limitations associated with the post-DFT methods, calculations using HSE06[12] and RPA were conducted with a smaller ($\sqrt{2} \times \sqrt{2}$)R45° cell, still restricting Brillouin-zone sampling to the $\Gamma$-point. We note that the finite-size effect was tested using DFT+U. For the 2H/Rh$_1$ (dihydride) configuration, the difference in adsorption energy between the two cell sizes was approximately 0.06 eV at the PBE+D3(BJ) level.

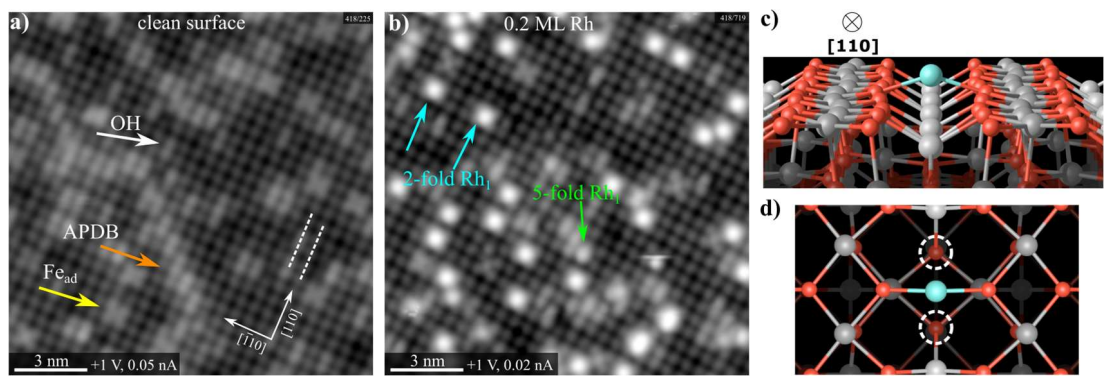

**Figure S1. Rh/Fe$_3$O$_4$(001) model catalyst structure.** (a) STM image of the clean Fe$_3$O$_4$(001) surface exhibiting the ($\sqrt{2}\times\sqrt{2}$)R45° reconstruction. The bright protrusions correspond to pairs of surface Fe atoms oriented along the [110] direction, thus forming Fe rows in this direction (white dashed lines).[1] Oxygen atoms are not visible in empty-state STM due to their low density of states near the Fermi level. Surface defects, including hydroxyl groups (OH), antiphase domain boundaries (APDB), and Fe adatoms (Fe$_{ad}$), are marked with white, orange, and yellow arrows, respectively. These defects have been characterized previously.[14] (b) STM image after deposition of 0.2 ML Rh at 300 K. Twofold oxygen-coordinated Rh adatoms appear between Fe rows (cyan arrows). Less common fivefold coordinated Rh adatoms are also observed (green arrow), as reported earlier.[15] (c, d) Perspective and top-view DFT-optimized structures of a twofold coordinated Rh$_1$ atom on the reconstructed Fe$_3$O$_4$(001) surface. The dashed circles in (d) indicate two equivalent subsurface oxygen atoms that can weakly interact with Rh$_1$ upon molecular adsorption.

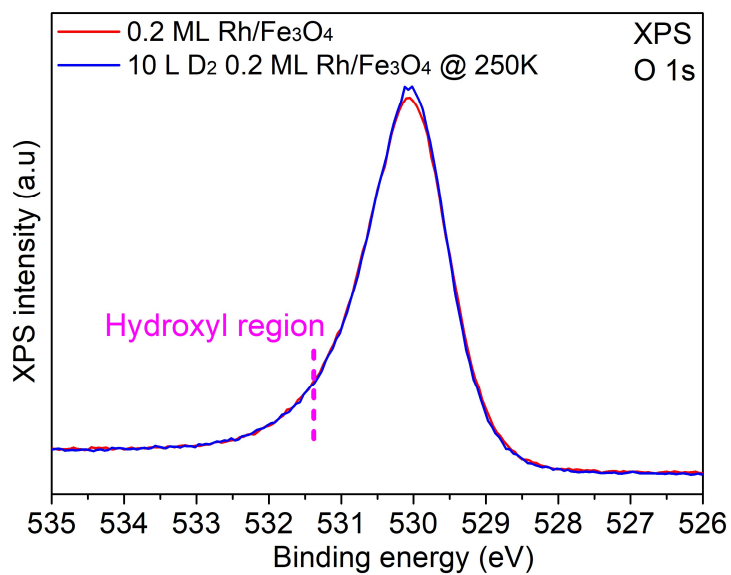

**Figure S2. XPS analysis confirms absence of OD formation after extended $D_2$ exposure.** O 1s XPS of 0.2 ML Rh/$Fe_3O_4$(001) are shown before and after exposure to 10 L $D_2$ at 250 K. The XPS was collected with a grazing angle of ≈ 71°. The pink dashed line marks the energy of OH (or OD) signals.[16] Even under these thermodynamically and kinetically favorable conditions (compared to 1 L $D_2$ at 200 K in Figure 2b), no OD-related signal is observed. This confirms that atomic hydrogen spillover onto the $Fe_3O_4$ support does not occur in this Rh single-atom catalyst system.

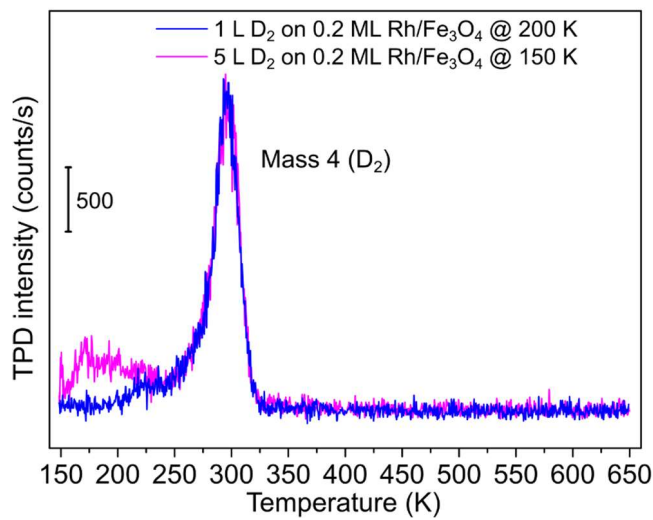

**Figure S3. Saturation behavior of D$_2$ adsorption on Rh/Fe$_3$O$_4$(001).** TPD spectra following adsorption of 1 L D$_2$ at 200 K (blue) and 5 L D$_2$ at 150 K on 0.2 ML Rh/Fe$_3$O$_4$(001). Despite differences in exposure temperature and dose, both spectra exhibit an overlapping desorption peak centered around 295 K, indicating that 1 L D$_2$ at 200 K is sufficient to saturate the Rh sites. The shoulder at 150–220 K is attributed to D$_2$ adsorption on Rh deposited onto the Ta sample plate. This contribution is further examined in the following figures.

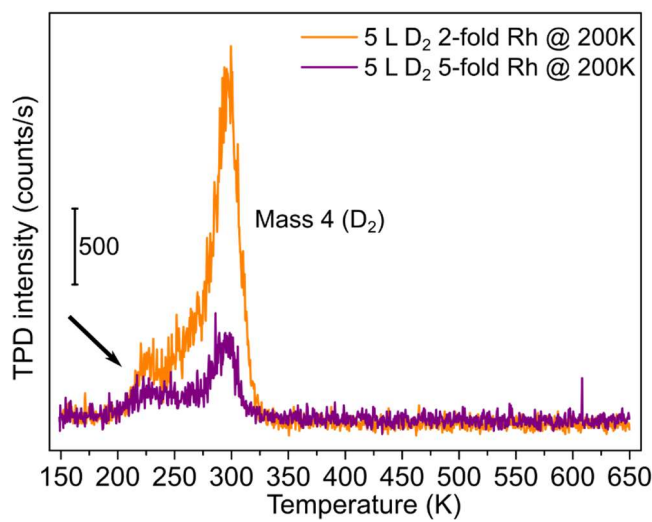

**Figure S4. Control experiment to assess the contribution of 5-fold coordinated Rh to $D_2$ desorption.** TPD spectra recorded after 5 L $D_2$ adsorption at 200 K on two Rh/Fe$_3$O$_4$(001) samples: one dominated by twofold coordinated Rh adatoms (orange) and another with a higher density of fivefold coordinated Rh adatoms embedded in the surface (purple).[15b] Both spectra exhibit a low-temperature shoulder (black arrow), but the signal intensity does not increase in the sample containing more fivefold Rh sites. This confirms that the shoulder feature does not originate from $D_2$ adsorption at fivefold Rh adatoms.

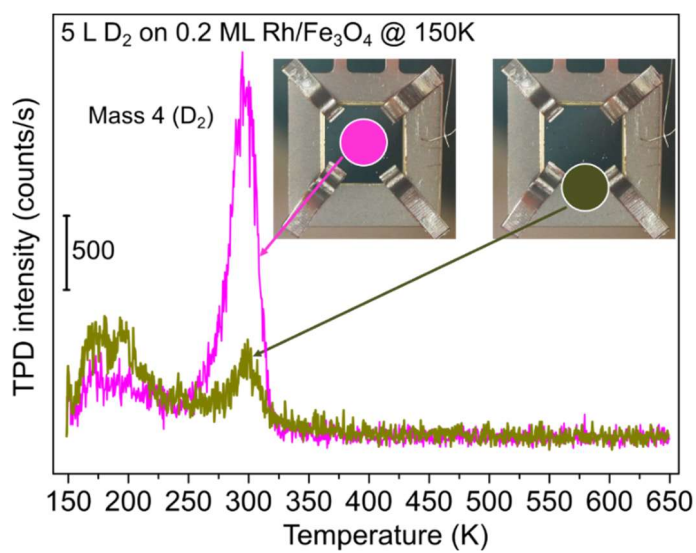

**Figure S5. Control experiment to assess $D_2$ adsorption on the Ta sample plate.** TPD spectra following 5 L $D_2$ exposure at 150 K on 0.2 ML Rh/Fe$_3$O$_4$(001). In the standard configuration (pink), $D_2$ is delivered via the molecular beam onto a 3.5 mm diameter area centered on the square sample. In the control experiment (olive green), the sample was displaced by 3.5 mm so that $D_2$ was directed primarily onto the Ta sample plate, with only a small contribution from on Rh/Fe$_3$O$_4$(001). The low-temperature desorption peak is stronger in the control spectrum, suggesting it originates from D$_2$ adsorbed on the Ta plate, not on the Fe$_3$O$_4$(001) surface.

Figures S4–S5 investigate the origin of the low-temperature shoulder or peak (marked by black arrow) at 150−220 K observed in the D$_2$ TPD spectra. Initially, we hypothesized that this feature might arise from D$_2$ adsorption on a minority population of fivefold coordinated Rh adatoms (Figure S1b, green arrow) formed during deposition.[15b] To test this, we prepared a surface enriched in fivefold Rh sites by annealing a 0.2 ML twofold Rh/Fe$_3$O$_4$(001) sample to 420 K, a treatment known to promote conversion from twofold Rh adatoms to embedded species, thereafter only a few twofold Rh sites remain.[17] As shown in Figure S4, this treatment does not increase the intensity of the shoulder peak, suggesting that fivefold Rh is not its origin.

We next considered whether the feature could arise from the Ta sample plate. Given the low sticking coefficient of D$_2$ on 0.2 ML Rh/Fe$_3$O$_4$(001) surface at 200 K, most of the dosed molecules are expected to scatter and be pumped away, but some may be scattered and adsorb on the surrounding sample mount. To test this, we intentionally displaced the sample by 3.5 mm so the molecular beam predominantly targeted the Ta plate. The resulting TPD trace (Figure S5, olive green) shows a pronounced low-temperature peak, suggesting that the feature originates from D$_2$ adsorption on the Ta substrate, not from Rh on Fe$_3$O$_4$(001).

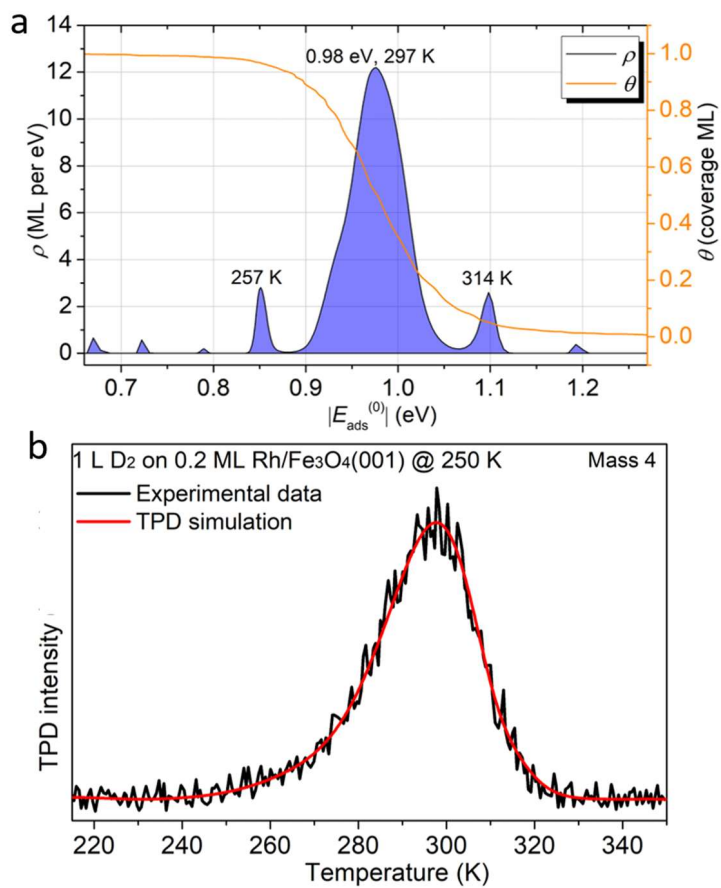

**Figure S6. Extraction and validation of adsorption energy from $D_2$ TPD data.** (a) Adsorption energy distribution $\rho(E_{ads}(0))$ extracted from the experimental TPD spectrum following 1 L $D_2$ exposure on 0.2 ML Rh/$Fe_3O_4$(001) at 250 K, using a recently developed TPD analysis framework.[18] The model assumes adsorption of $D_2$ on twofold coordinated $Rh_1$ sites. For the analysis, the density of adsorption sites (defined as 1 ML in this plot) was taken as the nominal Rh coverage, $2.82 \times 10^{13}$ cm$^{-2}$, the sticking probability was assumed as Langmuirian sticking with a zero-coverage sticking coefficient of 0.05, corresponding to a capture area of 18 Å² per unoccupied Rh atom (a rough guess). The analysis also depends on low-frequency vibrational modes, which contribute both to the energy and (more importantly) entropy. The DFT-calculated values are 10, 13 and 38 meV for Rh without an adsorbate, 29, 35 and 38 meV for Rh with two D atoms (dihydride), and the vibrational modes of the two adsorbed D atoms are at 36, 58, 62, 62, 192 and 193 meV. The gas-phase D–D stretch is frozen at the relevant temperature and can be neglected. Based on this analysis, the main desorption peak at 297 K corresponds to an adsorption energy of 0.98 eV. The small side peaks in (a) are caused by noise. Assuming an error by a factor of 10 in the sticking yields an adsorption energy errors of 0.06 eV. The errors induced by the uncertainty of the vibrational energies are negligible (5 meV for a change by 30%). (b) Simulated TPD trace generated from the extracted

energy distribution (red curves) overlaid on the experimental spectrum (black curve), showing good agreement between model and experiment.

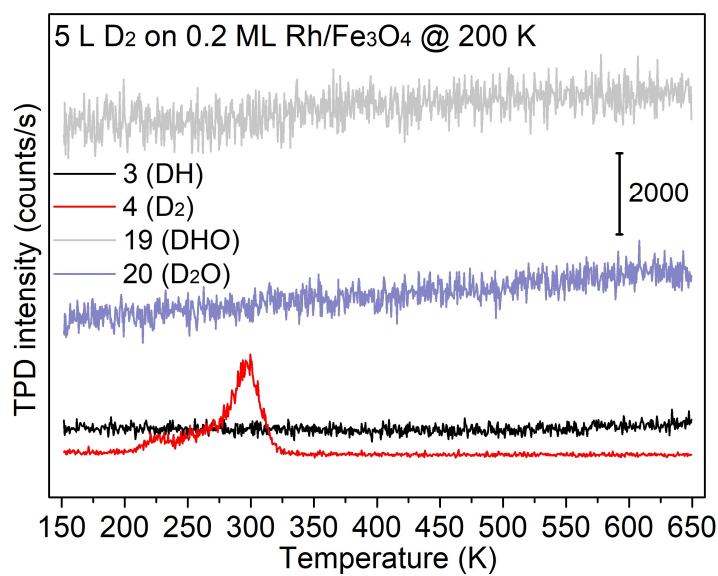

**Figure S7. TPD spectra with a high dose.** TPD spectra recorded following exposure of 0.2 ML Rh/Fe$_3$O$_4$(001) to 5 L D$_2$ at 200 K. This experiment was designed to test whether increased D$_2$ exposure promotes hydrogen spillover from isolated Rh sites to the Fe$_3$O$_4$ support, potentially leading to hydroxyl formation and subsequent water desorption during TPD. Compared to the 1 L D$_2$ experiment (Figure 3b), no water desorption is observed, indicating that spillover and formation of stable OD species do not occur even under extended D$_2$ exposure.

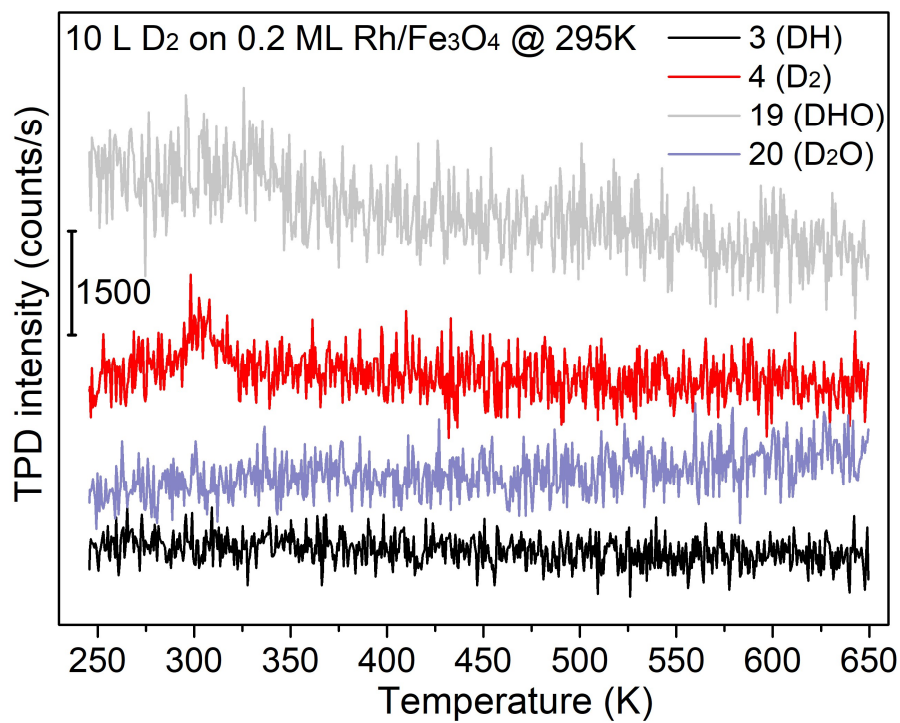

**Figure S8. TPD spectra following high-dose D₂ exposure at the near desorption temperature.** A 0.2 ML Rh/Fe₃O₄(001) sample was exposed to 10 L D$_2$ at 295 K, cooled to 250 K, and then subjected to TPD measurements from 250–650 K, monitoring masses 3, 4, 19, and 20. Only a weak D$_2$ desorption signal was detected near 300 K, indicating negligible adsorption at 295 K. The absence of water desorption further confirms that no D$_2$ dissociation or spillover occurs under these conditions.

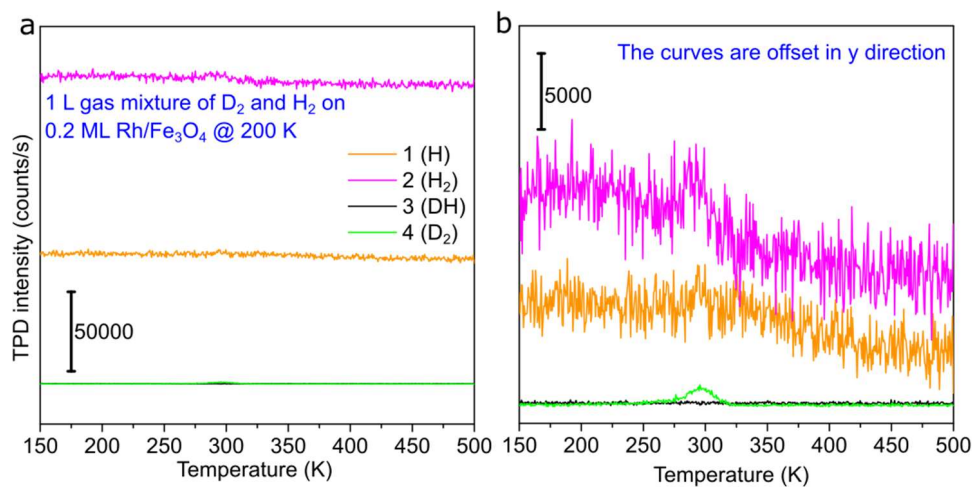

**Figure S9. H–D exchange experiment to test for hydrogen spillover.** A 1:1 mixture of $H_2$ and $D_2$ was dosed onto a 0.2 ML $Rh/Fe_3O_4$(001) sample at 200 K. TPD spectra monitoring masses 1, 2, 3, and 4 are shown in (a), with curves offset in (b) for clarity. Despite the presence of both $H_2$ and $D_2$ adsorption signals, no DH signal (mass 3) was detected. The high background of $H_2$ in UHV limits the signal-to-noise ratio for mass 1 and 2, which is why $D_2$ is preferred in surface science studies of hydrogen activation. This experiment is a test for the scenario of hydrogen spillover to the support, but desorption not in the form of water (which has been excluded in figures S7, S8) but via reverse spillover to the Rh, where it would recombine and desorb. In this scenario, isotopic scrambling, i.e., the formation of DH should be possible: H from $H_2$ adsorption at one Rh and D from $D_2$ adsorbed at a nearby Rh may desorb as DH; alternatively, DH could be formed by H and D from sequential adsorption events at the same Rh (each followed by spillover). The absence of DH (black trace) supports the conclusion that hydrogen activation is localized at $Rh_1$ sites without spillover onto the $Fe_3O_4$ support.

**Table S1.** DFT-calculated adsorption energies for dihydride and dihydrogen species on 2-fold coordinated $Rh_1/Fe_3O_4(001)$ using three different functionals: PBE+D3(BJ), r²SCAN, and HSE06, as well as an RPA calculation to better account for electronic correlation. RPA and HSE06 energies were corrected for zero-point vibrational contributions obtained from HSE06 frequencies. The cell size was $(2\sqrt{2} \times 2\sqrt{2})R45°$ for PBE+D3(BJ) and r²SCAN. For the computationally more demanding HSE06 and RPA calculations, a $(\sqrt{2} \times \sqrt{2})R45°$ cell was used. The $E_{ads}$ difference for $2H/Rh_1$ (dihydride) between these cell sizes is ≈0.06 eV at PBE+D3(BJ) level.

|  | PBE + D3(BJ) | | r²SCAN | | HSE06 | | RPA | |
|---|---|---|---|---|---|---|---|---|
|  | $\Delta E_{ads}$ (eV) | $\Delta E_{ads,(ZPE)}$ (eV) | $\Delta E_{ads}$ (eV) | $\Delta E_{ads,(ZPE)}$ (eV) | $\Delta E_{ads}$ (eV) | $\Delta E_{ads,(ZPE)}$ (eV) | $\Delta E_{ads}$ (eV) | $\Delta E_{ads,(ZPE)}$ (eV) |
| $2H/Rh_1$ Dihydride | −1.31 | −1.11 | −1.04 | −0.86 | −1.47 | −1.28 | −1.33 | −1.14 |
| $H_2/Rh_1$ Dihydrogen | −1.08 | −0.89 | −0.98 | −0.81 | −1.14 | −0.97 | −1.15 | −0.98 |
| $2D/Rh_1$ Dihydride | −1.31 | −1.16 | −1.04 | −0.90 | −1.47 | −1.33 | −1.33 | −1.19 |
| $D_2/Rh_1$ Dihydrogen | −1.08 | −0.94 | −0.98 | −0.85 | −1.14 | −1.02 | −1.15 | −1.03 |

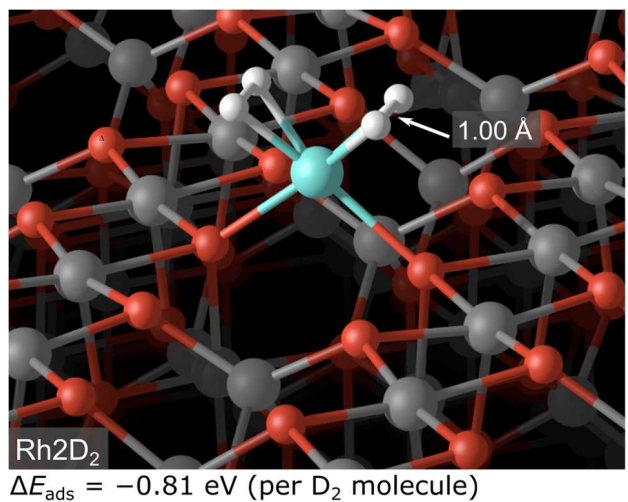

**Figure S10. Simultaneous adsorption of two D₂ molecules on 2-fold coordinated Rh₁/Fe₃O₄ (001).** DFT calculations show that two D₂ molecules can coordinate to a single Rh₁ site in the form of dihydrogen, with a total adsorption energy of −1.62 eV (PBE+D3(BJ)).

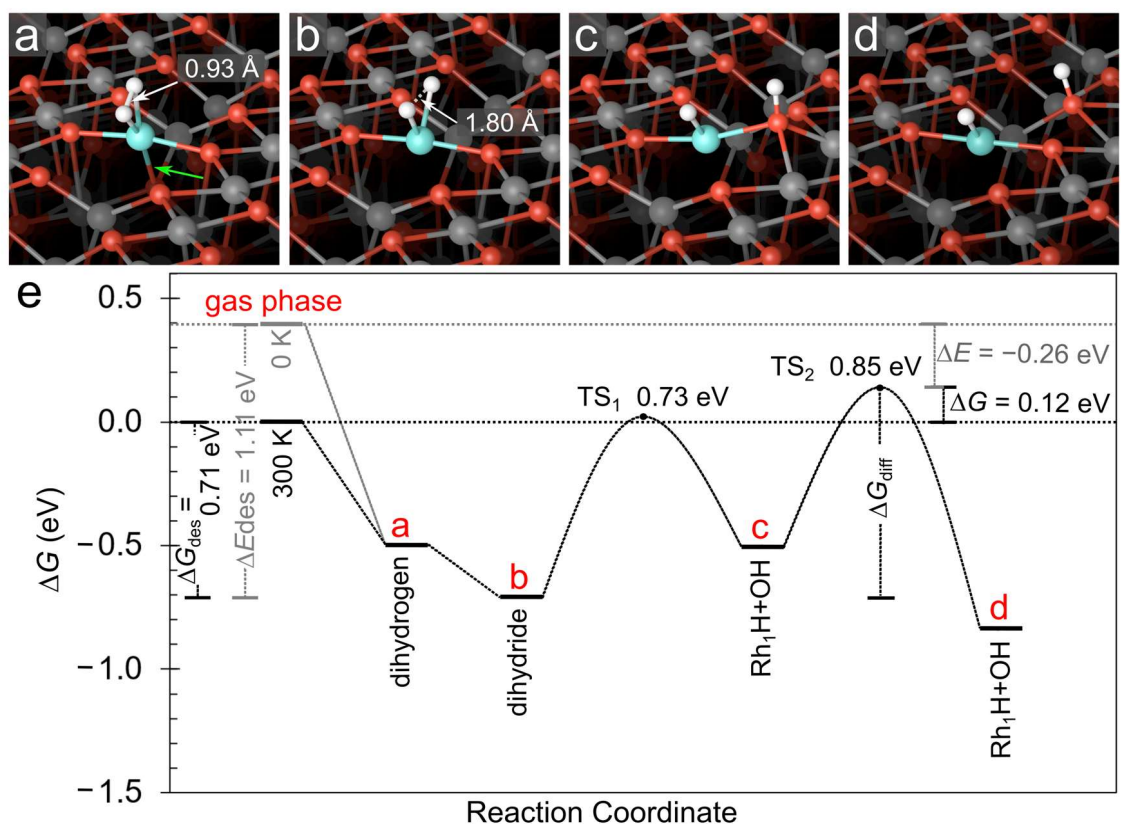

**Figure S11. DFT-calculated mechanism of H$_2$ dissociation and spillover on Rh$_1$/Fe$_3$O$_4$(001).** Perspective views are shown for (a) a dihydrogen configuration and (b) a dihydride configuration. In both cases, the Rh$_1$ species is bond to two surface oxygen atoms. An additional weak bond from the Rh to a subsurface O atom (roughly 0.3 Å shorter than in the Rh2H configuration) is indicated by the green arrow. (c-d) Atomic structures of atomic H spillover along the Fe$_3$O$_4$(001) surface. Fe atoms are shown in dark gray, and oxygen atoms are red. (e) Computed reaction energy profile for molecular H$_2$ adsorption, dissociation, and hydrogen spillover from a Rh$_1$ site onto the Fe$_3$O$_4$(001) substrate. The reported Gibbs free energy ($\Delta G$) values incorporate the entropy difference between the adsorbed state ($S_{ads} = 0$) and the gas phase ($S_{gas}$) (as referenced in database).[8a] These $\Delta G$ values are presented for the experimental desorption temperature (300 K) and derived from DFT calculations that include ZPE corrections. The overall reaction mechanism remains identical to the 0 K pathway, except the desorption step, for which the 0 K desorption energy ($\Delta E_{ads}$) is provided for comparative reference. Transition states (TS$_1$ and TS$_2$) and the corresponding activation barriers with respect to the dihydride state (b) are indicated.

**Table S2**. DFT-calculated vibration frequencies of dihydride adsorption on Rh/Fe$_3$O$_4$(001). The first two values are Rh–H(D) stretch, the next four are H(D) wagging, and the last three are Rh modes.

| 2H/Rh/Fe$_3$O$_4$(001) | 275.8 meV, 272.6 meV, 89.5 meV, 88.4 meV, 83.8 meV, 53.1 meV, 34.5 meV, 16.6 meV, 14.5 meV |
|---|---|
| 2D/Rh/Fe$_3$O$_4$(001) | 196.1 meV, 193.4 meV, 64.3 meV, 62.7 meV, 59.6 meV, 38.5 meV, 33.6 meV, 16.4 meV, 14.2 meV |